\documentclass[twocolumn, journal]{IEEEtran}

\IEEEoverridecommandlockouts
\usepackage[table]{xcolor}
\usepackage{color}
\usepackage{graphicx}
\usepackage{amsmath}
\usepackage{amssymb}
\usepackage{algorithm}
\usepackage{algorithmic}
\usepackage{amsmath}
\usepackage{multirow}
\usepackage{booktabs}
\usepackage{array}
\usepackage{amsthm}
\usepackage{stfloats}
\usepackage{caption}
\usepackage{subfigure}
\usepackage{bm}
\usepackage{setspace}
\usepackage{chngpage}
\usepackage{pifont}

\newcommand{\be}{\begin{equation}}
\newcommand{\ee}{\end{equation}}
\newcommand{\bea}{\begin{eqnarray}}
\newcommand{\eea}{\end{eqnarray}}
\newcommand{\ba}{\begin{array}}
\newcommand{\ea}{\end{array}}

\allowdisplaybreaks[4]

\title{Integrated Sensing and Communication with Reconfigurable Intelligent Surfaces:  Opportunities, Applications, and Future Directions
\thanks{R. Liu, M. Li, and H. Luo are with the School of Information and Communication Engineering, Dalian University of Technology, Dalian 116024, China (e-mail: liurang@mail.dlut.edu.cn; mli@dlut.edu.cn; luohonghao@mail.dlut.edu.cn).}
\thanks{Q. Liu is with the School of Computer Science and Technology, Dalian University of Technology, Dalian 116024, China (e-mail: qianliu@dlut.edu.cn).}
\thanks{A. L. Swindlehurst is with the center for Pervasive Communications and Computing, University of California, Irvine, CA 92697, USA (e-mail:
swindle@uci.edu).}
}
\author{Rang Liu,~\IEEEmembership{Graduate Student Member,~IEEE,}
        Ming Li,~\IEEEmembership{Senior Member,~IEEE,}
        Honghao Luo,\\
        Qian Liu,~\IEEEmembership{Member,~IEEE,}
        and A. Lee Swindlehurst,~\IEEEmembership{Fellow,~IEEE}
        }
\begin{document}
\maketitle
\thispagestyle{empty}
\begin{abstract}
Integrated sensing and communication (ISAC) is emerging as a key enabler to address the growing spectrum congestion problem and satisfy increasing demands for ubiquitous sensing and communication.
By sharing various resources and information, ISAC achieves much higher spectral, energy, hardware, and economic efficiencies.
Concurrently, reconfigurable intelligent surface (RIS) technology has been deemed as a promising approach due to its capability of intelligently manipulating the wireless propagation environment in an energy and hardware efficient manner.
In this article, we analyze the potential of deploying RIS to improve communication and sensing performance in ISAC systems.
We first describe the fundamentals of RIS and its applications in traditional communication and sensing systems, then introduce the principles of ISAC and overview existing explorations on RIS-assisted ISAC, followed by one case study to verify the advantages of deploying RIS in ISAC systems.
Finally, open challenges and research directions are discussed to stimulate this line of research and pave the way for practical applications.
\end{abstract}


\section{Introduction}

While intensive commercialization of the fifth generation mobile communication system (5G) has been pushed forward worldwide, research on the sixth generation (6G) has been launched toward the goal of smart connectivity of everything.
To support the visions of 6G for typical applications such as smart city/transportation/home, where numerous wireless devices are deployed to sense the environment and/or communicate with each other, it is expected that the communication, sensing, and computational functions of the network will converge, calling for revolutionary technologies and radical changes to existing platforms.
Integrated sensing and communication (ISAC) is a key enabling technology \cite{FLiu 2021}, \cite{Zhang ICST 2022} that allows radar sensing and wireless communications to share various resources including spectrum, hardware architecture, signal processing platform, etc.
In addition to leveraging congested resources, ISAC encourages mutual assistance by sharing information to improve communication and sensing functionalities, contributing to improved performance in terms of spectral/energy/hardware/cost efficiency.

The integration of sensing and communication systems is facilitated by the fact that they employ similar radio frequency (RF) front-end hardware, a similar signal processing algorithm framework, a consistent evolution towards wideband systems with massive antennas, etc.
To enjoy the integration and coordination gains in ISAC systems, the inherently conflicting requirements of communication and sensing should be carefully accounted for in order to achieve satisfactory trade-offs.
Towards this goal, multi-input multi-output (MIMO) antenna array architectures have been employed to exploit spatial degrees of freedom (DoFs).
Nevertheless, deteriorated performance in uncontrollable electromagnetic environments is still inevitable, which motivates use of the emerging reconfigurable intelligent surface (RIS) technology \cite{Renzo JSAC 2020}.

RIS has been deemed as an important enabling technology for future 6G networks owing to its ability to reshape electromagnetic environments in an energy/hardware-efficient manner.
The deployment of RIS introduces additional optimization DoFs and can establish virtual  line-of-sight (LoS) links for users in single-blind or shadowed areas or at the cell edges.
In light of these benefits, the application of RIS to various wireless communication scenarios has been extensively investigated.
While the advantages of deploying RIS in communication systems have already been well established, research in the context of radar and especially ISAC has only recently begun.
Therefore, in this article we explore the potential of RIS in enhancing radar sensing functions and discuss the opportunities, applications, and future directions for RIS-assisted ISAC.

\section{RIS-Assisted Communications/Sensing}

Already introduced and validated for communication systems, RIS technology has attracted considerable attention from researchers in widespread fields.
Given the already large number of works on RIS-assisted communications, in this section we briefly introduce RIS-assisted communications, and elaborate on the applications of RIS to sensing systems.

\subsection{Fundamentals of RIS \& RIS-Assisted Communications}

An RIS is generally a two-dimensional planar array consisting of large numbers of passive reflecting elements, which have simple hardware structures and low energy consumption and cost.
The electromagnetic properties of the incident signals, e.g., phase-shift, amplitude, frequency, polarization, etc., can be independently tuned by controlling the parameters of passive electronic circuits attached to each element.
Cooperatively adjusting these reflecting coefficients offers considerable passive beamforming gains that can be used to improve various system performance metrics such as maximizing spectral efficiency/sum-rate/energy efficiency/secrecy rate/received energy, minimizing transmit power/symbol error rate, etc.
Furthermore, the lightweight and thin construction of these devices provides the RIS with significant flexibility, enabling it to be easily deployed on the surfaces of tall buildings or other portable and mobile devices.
Widespread applications have revealed the advantages of deploying RIS in wireless communication systems.

\subsection{RIS-Assisted Sensing}

As research on RIS-assisted wireless communications is steadily proceeding, explorations of deploying RIS in sensing systems have also emerged.
Before discussing existing investigations on RIS-assisted sensing, we begin with some basics of sensing systems.
In radar sensing, target detection and parameter estimation are two primary tasks.
Specifically, target detection refers to the process of identifying the presence or absence of a target in a reflected radar return, and performance is often evaluated by the probabilities of detection and false alarm.
Estimation performance for parameters such as target azimuth angle/distance/velocity is typically evaluated by the mean squared error (MSE) or the Cram\'{e}r-Rao lower bound (CRLB).
Considering the difficulties of directly optimizing these metrics, however, most existing works use alternative optimization criteria such as the illuminated signal power, the transmit beampattern, the signal-to-noise ratio (SNR) or signal-to-interference-plus-noise ratio (SINR) of the received echo signals, etc.

The deployment of RIS opens up new opportunities for traditional sensing applications.
By creating an effective LoS link and providing additional DoFs for optimization, deploying an RIS can boost the desired target returns and suppress interference by manipulating the propagation environment, which not only enhances the sensing performance for targets that already enjoys LoS propagation, but also allows the radar to sense targets in shadowed areas that would normally be invisible to the radar.
In the following, we will review examples of investigations into RIS-assisted sensing based on different sensing metrics and goals.

\subsubsection{CRLB}

An RIS-enabled sensing system was investigated based on the CRLB metric in \cite{Song arXiv 2022}.
Assuming the direct LoS link is blocked, it is shown that the deployment of an RIS enables the access point (AP) to estimate the target's direction of arrival (DoA) by processing the echo signals from the virtual LoS link.
The CRLB for DoA estimation was maximized by jointly optimizing the transmit beamforming and RIS reflecting coefficients.

\subsubsection{SNR}

In \cite{Buzzi 2021}, RISs are deployed near the radar transmitter and receiver to facilitate steering transmit signals toward the target and collecting target returns for the radar receiver.
By using orthogonal transmit waveforms and the generalized likelihood ratio test (GLRT) for target detection, the RIS reflecting coefficients were designed to maximize the SNR of the received echo signals to achieve a higher detection probability.
In addition, the RIS is exploited to boost the received signal power at desired locations while suppressing the signals at undesired locations in  \cite{FWang 2021}.
To fully exploit the DoFs introduced by multiple transmit antennas, the transmit beamforming of the AP and the RIS reflecting coefficients were jointly optimized.

\subsubsection{DoA estimation}

An RIS-assisted unmanned aerial vehicle (UAV) swarm system was proposed in \cite{Chen TSP 2022}.
The UAV swarm mounted with RISs assists the central UAV in DoA estimation, which not only significantly reduces implementation cost, but also enables the central UAV to ``see'' the targets from multiple angles.
For achieving better DoA estimation performance, an atomic norm-based method was further developed.

\subsubsection{Localization and mapping}

Deploying RIS can also efficiently improve localization and mapping accuracy, resolution, and coverage.
By intelligently manipulating the propagation environment, the differences in signals reflected from different locations or targets can be enlarged to facilitate the receiver's ability to localize and create an accurate radio map.


\begin{table*}[htb]\small\begin{center}
\begin{tabular}{>{\columncolor{teal!20}}m{0.5cm}  >{\columncolor{gray!20}}m{2.3cm}  >{\columncolor{teal!20}}m{2.3cm} >{\columncolor{gray!20}}m{2.1cm}  >{\columncolor{teal!20}}m{2.1cm} >{\centering\columncolor{gray!20}}m{3.4cm} >{\columncolor{teal!20}}m{2.6cm} }
Ref.    &System model &Communication \& sensing metrics &Role of RIS & Echo paths &Optimization variables & Algorithms \\
\hline
\cite{He JSAC 2022}  & RCC, SU, MT & SINR; SINR & aid comm/radar & only LoS  &radar transmit and receive beamforming &PDD, CCCP \\
\specialrule{0em}{.2pt}{.2pt}
\cite{Sankar SPAWC 2021}      & DFRC, SU, ST & SINR; SINR & aid comm/radar & only NLoS  &only RIS reflecting coefficients & codebook\\
\specialrule{0em}{.2pt}{.2pt}
\cite{Wang TVT 2021b}     & DFRC, MU, MT  & MUI; CRLB & aid comm &only LoS & transmit waveform & exact penalty, manifold optimization\\
\specialrule{0em}{.2pt}{.2pt}
\cite{Wei JCS 2022}       & wideband DFRC, MU, ST, clutter  & SINR; SINR &aid both& only NLoS &transmit beamforming & Dinkelbach, ADMM, MM\\
\specialrule{0em}{.2pt}{.2pt}
\cite{Yan JCS 2022}       & DFRC, SU, ST & SNR; SNR &aid both &LoS\&NLoS &transmit beamforming & MM, SDR\\
\specialrule{0em}{.2pt}{.2pt}
\cite{Rang EUSIPCO 2022}  & DFRC, MU, ST   & sum-rate; SNR &aid both &LoS\&NLoS &transmit beamforming and receive filter & FP, MM, ADMM\\
\specialrule{0em}{.2pt}{.2pt}
\cite{Rang JSTSP 2022}    & DFRC, MU, ST, clutter   & Euclidean distance; SINR &aid both&LoS\&NLoS &transmit waveform and receive filter & MM, ADMM\\
\specialrule{0em}{.2pt}{.2pt}
\cite{Mishra ICASSP 2022}     &DFRC, MU, MT, PLS  & secrecy rate; SNR &aid comm/both &only LoS/NLoS&transmit beamforming, AN &MM
\end{tabular}\end{center}
\begin{small}\caption{Summary of research on RIS-assisted ISAC (SU: single user, MU: multi-user, ST: single target, MT: multi-target, the optimization variables naturally include the RIS reflecting coefficients).}\end{small}\vspace{0.2 cm}
\end{table*}

\section{Applications of RIS in ISAC Systems}

Exploiting the deployment of RIS in stand-alone communication or sensing systems, researchers have begun to explore its applications in ISAC systems.
In this section, we first introduce the principles of ISAC, then discuss some initial investigations for RIS-assisted ISAC systems, and finally present one case study to demonstrate the advantages of deploying RIS in such systems.

\subsection{Principles of ISAC}

While wireless communication and radar sensing have been evolving in parallel with limited interplay for decades, ISAC is recently emerging as an enabling technology to support the demands for high-quality ubiquitous wireless connectivity and high-accuracy sensing capabilities in future 6G networks.
ISAC permits different levels of integration, of which radar-communication coexistence (RCC) and dual-functional radar-communication (DFRC) are two main applications.
While RCC refers to simply sharing spectrum for co-existing radar and communication systems, DFRC exploits a fully-shared hardware platform and a unified transmit waveform to simultaneously perform communication and sensing functions.
Together with the advantages of shared resources, it is also very challenging to achieve a satisfactory trade-off between sensing and communication functions for both RCC and DFRC systems.

RCC relies on sophisticated interference management and cooperation between the radar and communication transmitters, which requires side-information exchange and thus results in high overhead, but minimal modifications to existing system hardware is needed.
While DFRC naturally achieves full cooperation using a single platform, the inherently conflicting requirements of sensing and communication functions pose challenges for dual-functional waveform designs.
For both RCC and DFRC systems, the transmit beamforming/waveform design is crucial to enhance the sensing and communication functions, and thus has been widely investigated under different metrics in the literature.
Research on ISAC is on the rise, and there is great potential for interplay with other advanced technologies towards better performance.

\begin{figure*}[!t]
\centering
\subfigure[]{
\begin{minipage}{0.32\linewidth}
\centering
\includegraphics[width = 1\linewidth]{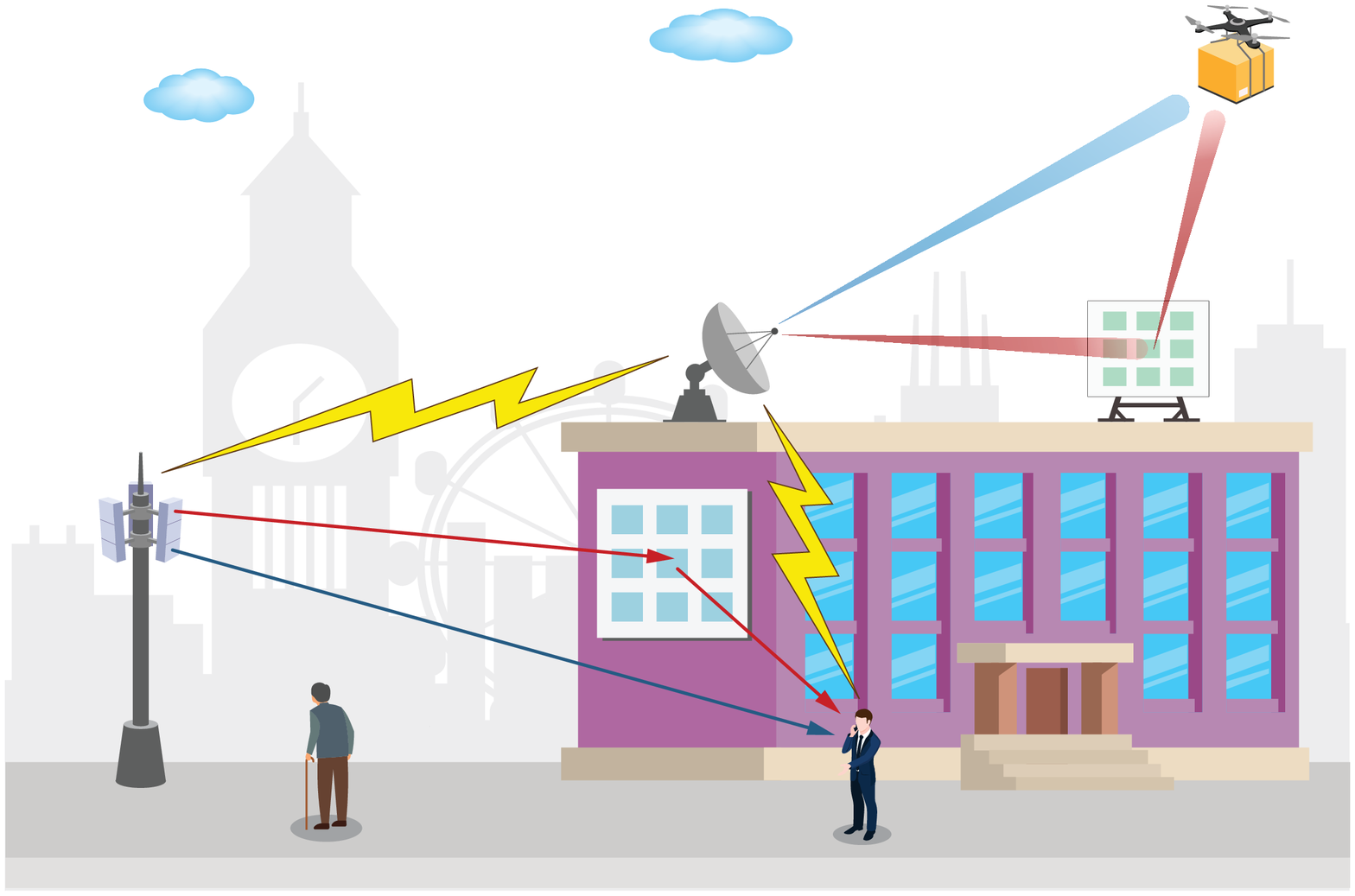}\vspace{0.2 cm}
\end{minipage}}
\subfigure[]{
\begin{minipage}{0.32\linewidth}
\centering
\includegraphics[width = 1\linewidth]{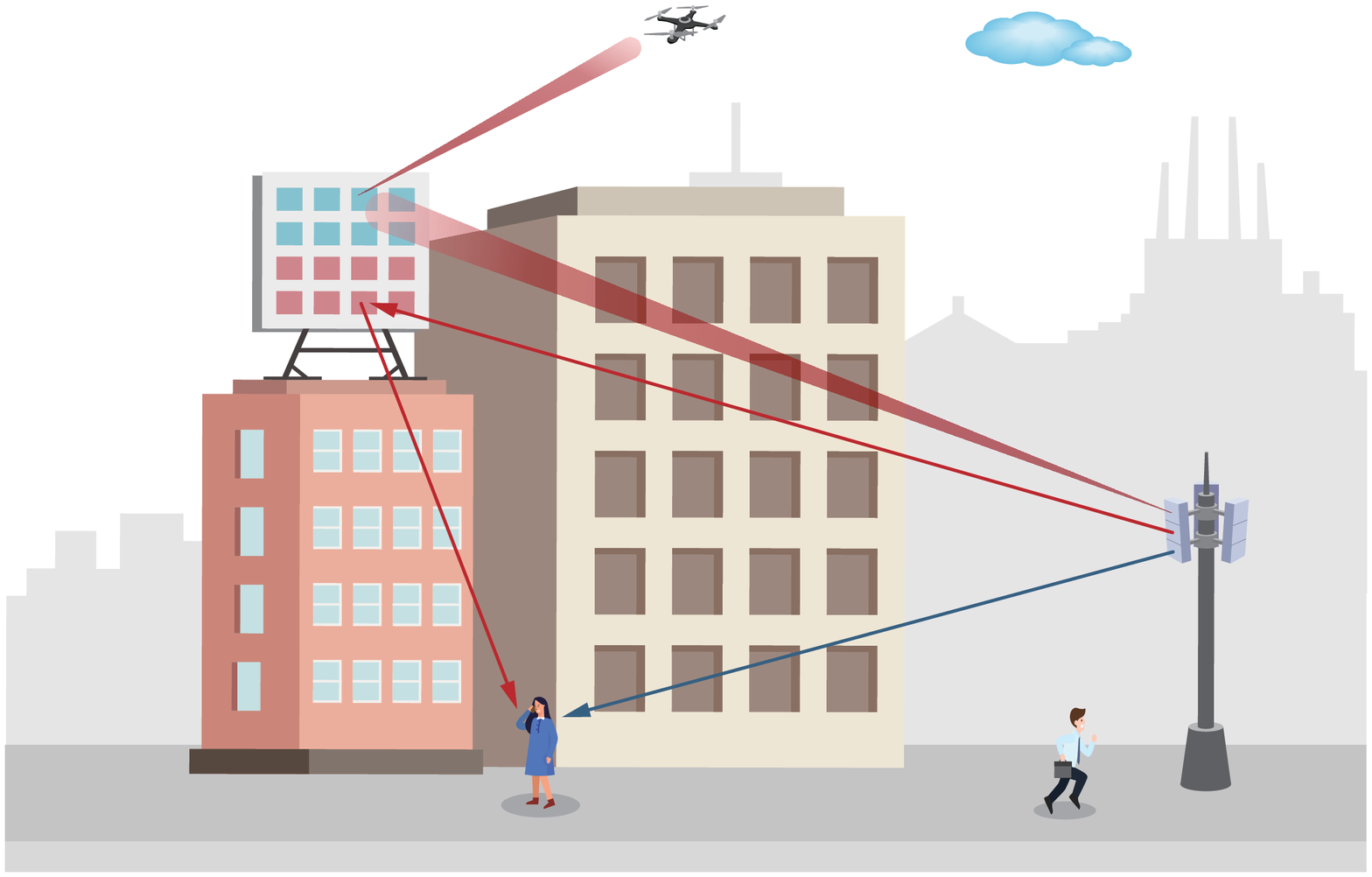}\vspace{0.2 cm}
\end{minipage}}
\subfigure[]{
\begin{minipage}{0.32\linewidth}
\centering
\includegraphics[width = 1\linewidth]{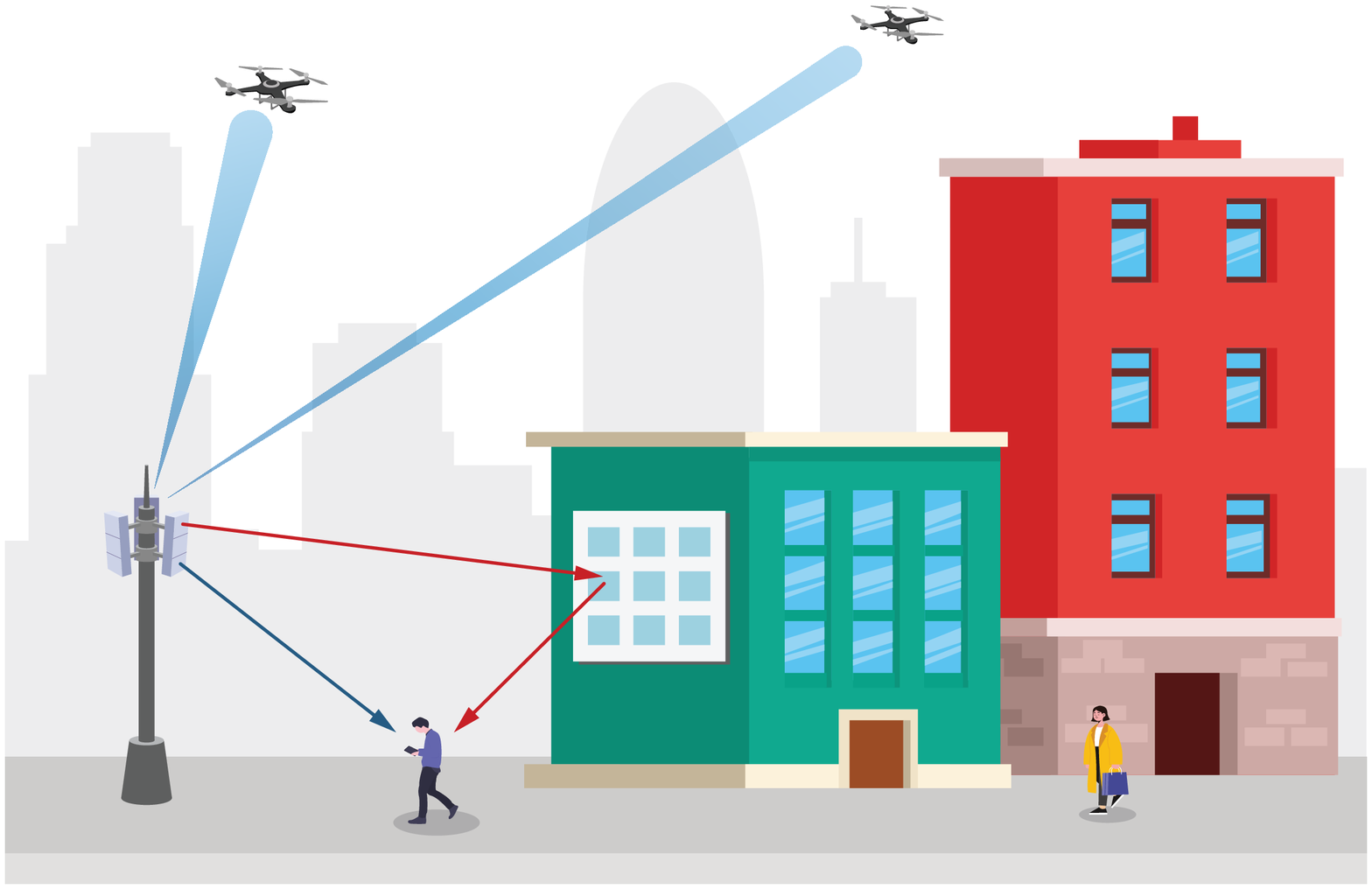}\vspace{0.2 cm}
\end{minipage}}
\subfigure[]{
\begin{minipage}{0.32\linewidth}
\centering
\includegraphics[width = \linewidth]{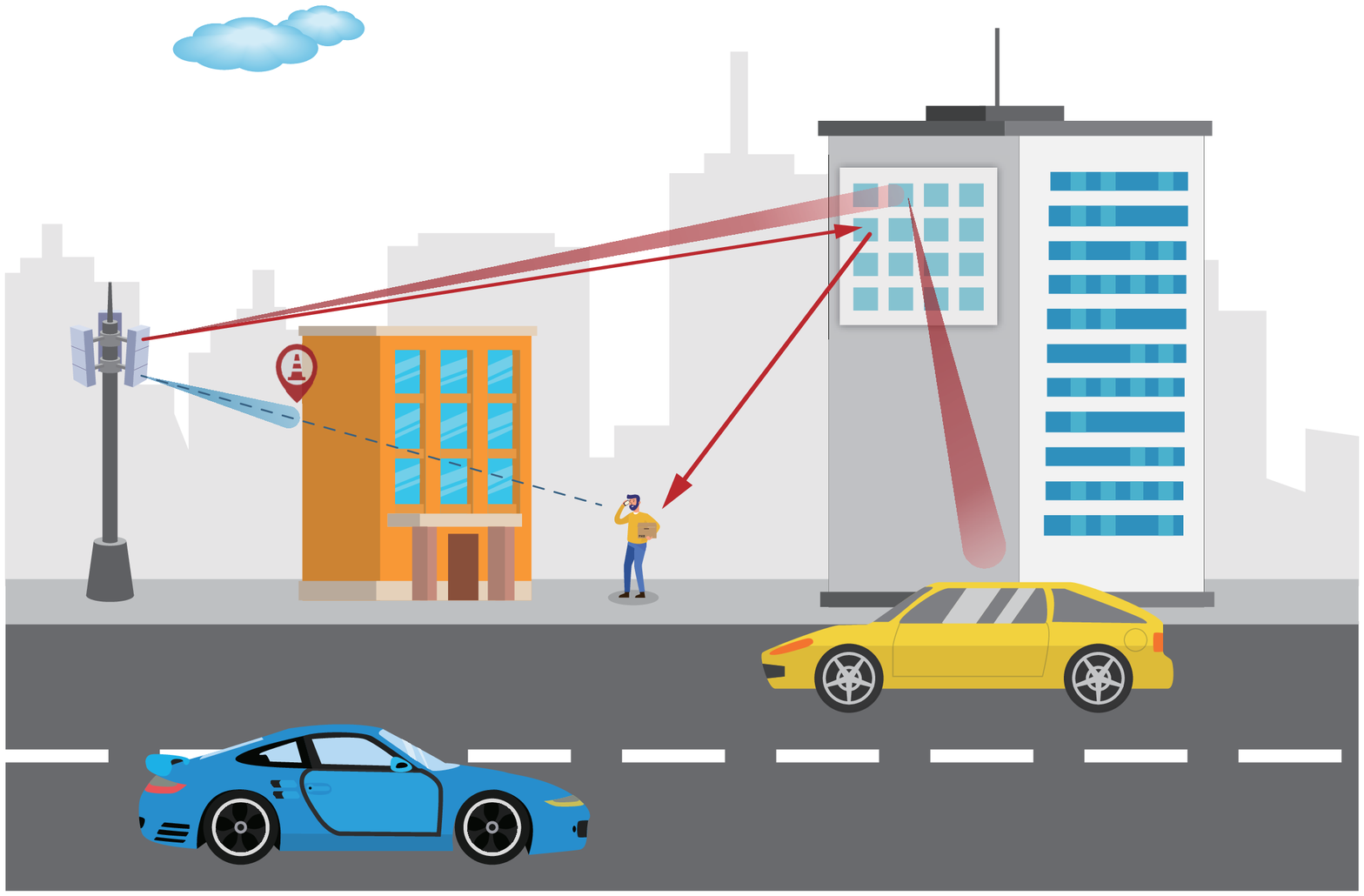}\vspace{0.2 cm}
\end{minipage}}
\subfigure[]{
\begin{minipage}{0.32\linewidth}
\centering
\includegraphics[width = \linewidth]{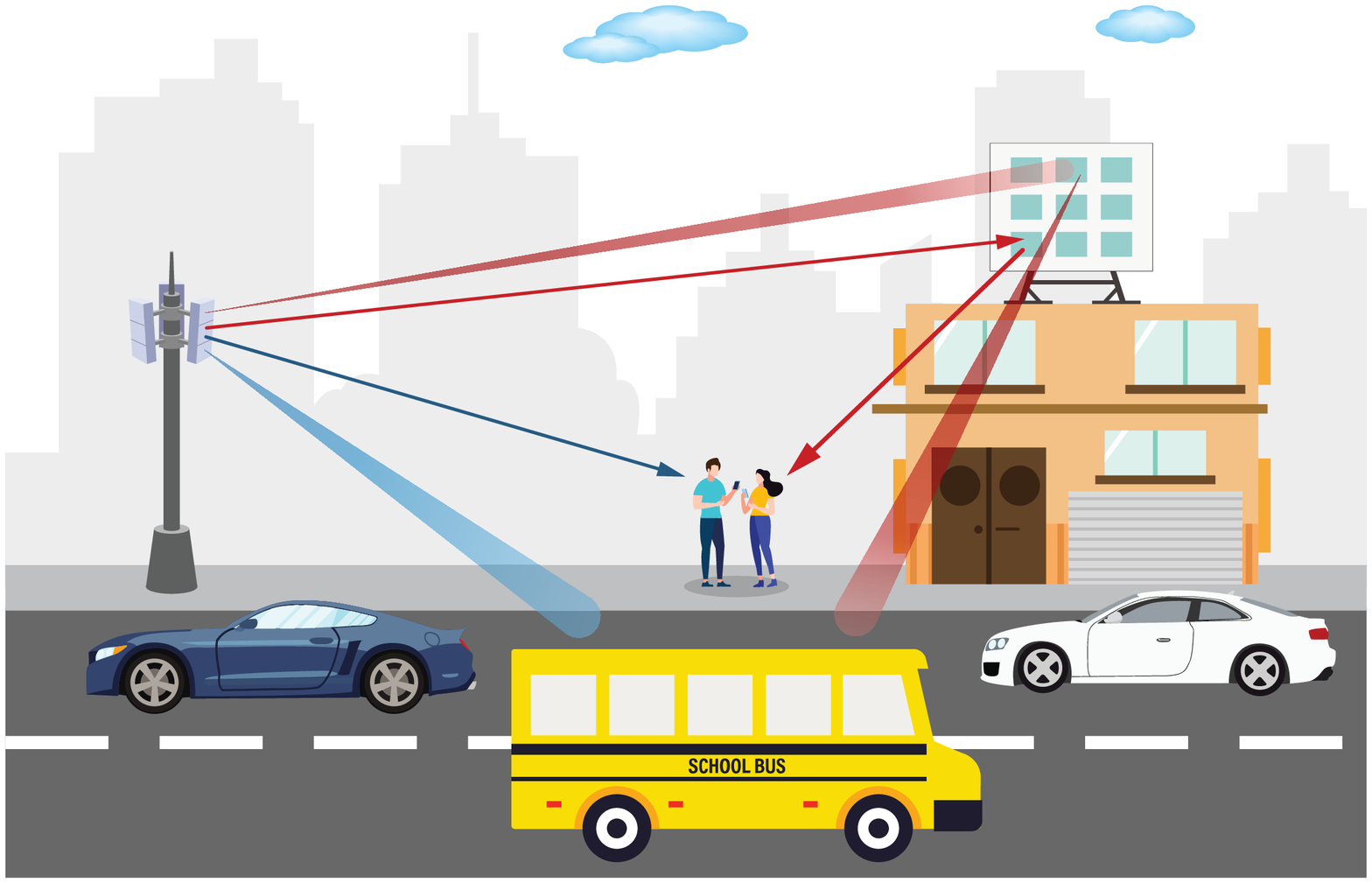}\vspace{0.2 cm}
\end{minipage}}
\subfigure[]{
\begin{minipage}{0.32\linewidth}
\centering
\includegraphics[width = \linewidth]{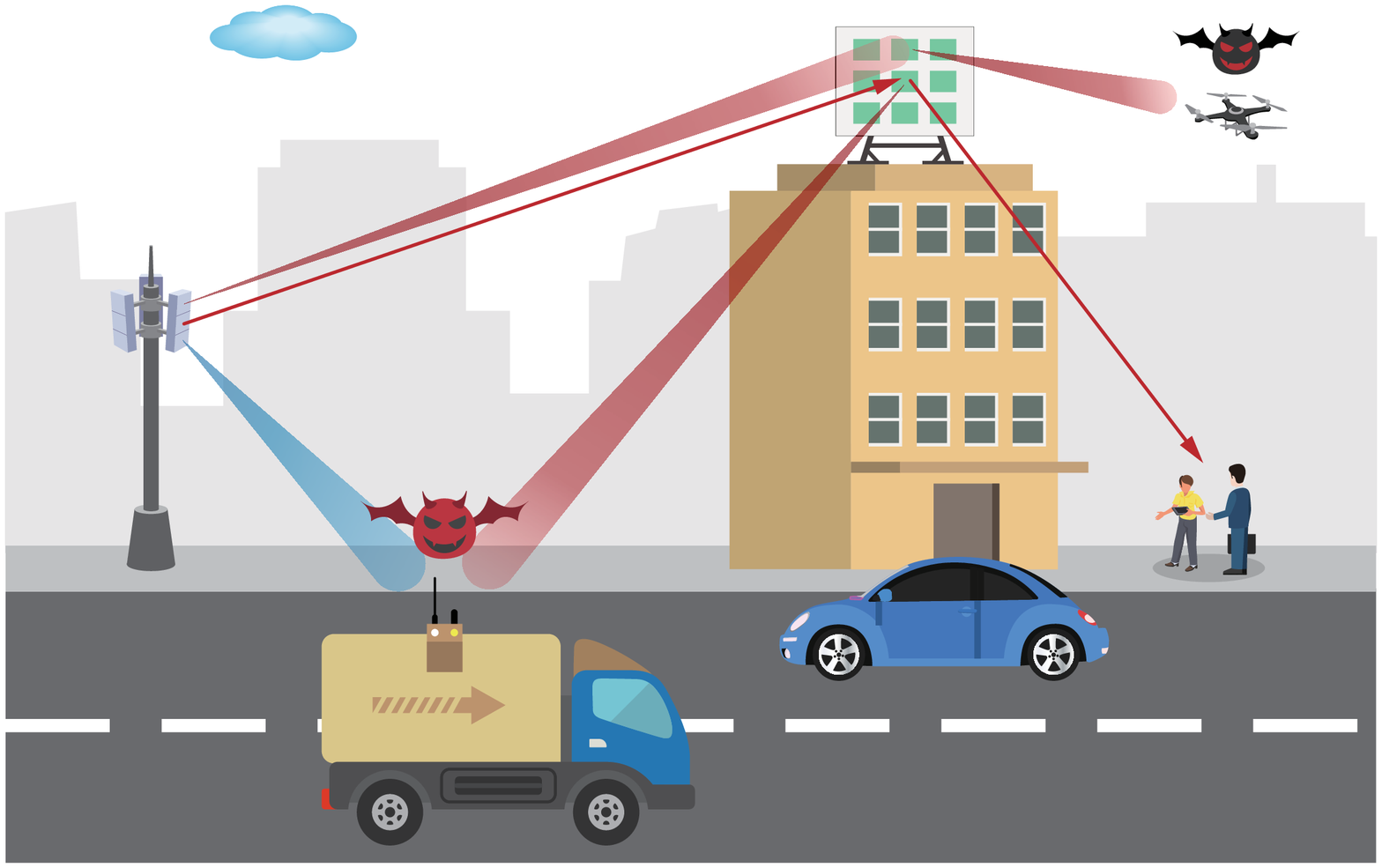}\vspace{0.2 cm}
\end{minipage}}\vspace{0.1 cm}
\begin{small}\caption{Typical applications of RIS in ISAC systems: (a) RIS aids interference management in a RCC system. (b) Adaptively partition the RIS elements for assisting sensing and communication functions in a DFRC system. (c) RIS only assists communications in a DFRC system where the LoS link between the BS and the target is relatively strong. (d) RIS assists both sensing and communication in a DFRC system where the direct paths between the BS and the targets/users are blocked. (e) A more general RIS-assisted DFRC system where both the LoS and NLoS links have a non-negligible impact on sensing and communication. (f) An RIS-assisted DFRC system where the target is a potential eavesdropper.}\end{small}
\label{fig:RIS-assisted ISAC systems}\vspace{0.2 cm}
\end{figure*}

\subsection{RIS-assisted ISAC systems}

In light of above descriptions, we can foresee that the flexibility introduced by RIS will not only enable better communication and sensing performance, but also novel application scenarios, which in turn call for accurate models and the design of associated algorithms.
State-of-the-art research on RIS-assisted ISAC in the literature is summarized in Table I, and the corresponding applications are illustrated in Fig. 1.
We will elaborate on them in the following.

\subsubsection{Deployment in RCC systems}

RIS creates an opportunity to address the critical interference management issue in RCC systems.
By intelligently manipulating the propagation environment, deploying RISs can boost desired signals and suppress undesired signals to offer satisfactory sensing and communication performance in the presence of mutual interference, as shown in Fig. 1(a).
In addition, considering that the base station (BS) is usually located outside the guard zone of the radar to protect it from undesired interference, the deployment of RIS enables coverage for the users in the radar exclusion zone and improvement of the communication quality-of-service (QoS).
Another specific RIS-assisted RCC system was investigated in \cite{He JSAC 2022}, in which one RIS is placed near the BS to suppress the interference from it to the radar and another RIS near the user to cancel the interference from the radar.
The transmit beamforming of the radar and the RIS reflecting coefficients were jointly optimized to maximize the communication SINR under a radar SINR constraint for target detection and a total power constraint.
An alternating algorithm based on the penalty dual decomposition (PDD) and concave-convex procedure (CCCP) was developed to solve the resulting non-convex optimization problem.

\subsubsection{Deployment in DFRC systems}

In DFRC systems, RIS provides additional DoFs for the dual-functional waveform design, leading to enhanced trade-offs between communication and sensing performance for diverse applications.
Meanwhile, the large number of coupled variables lead to complicated non-convex optimization problems that require sophisticated optimization algorithms.
Following extensive research on DFRC, researchers have conducted studies on RIS-assisted DFRC for various RIS deployment scenarios as described below.

In order to achieve a low-complexity design, the authors in \cite{Sankar SPAWC 2021} proposed to adaptively partition the RIS reflecting elements to separately perform communication and sensing functionalities as shown in Fig. 1(b).
A multi-stage hierarchical codebook was then designed to localize the target while ensuring a strong communication link to the user.

As shown in Fig. 1(c), the RIS can be deployed to only assist communication users under the assumption that the signal directly reflected by the targets dominates the power of the echo signals.
This assumption is valid for the scenario of detecting targets located in open areas, where the LoS link between the BS and the target is relatively strong while that between the BS and the RIS is very weak.
In this scenario, the authors in \cite{Wang TVT 2021b} proposed to minimize the multi-user interference (MUI) under a CRLB constraint for DoA estimation, a constant-modulus waveform constraint, and the discrete RIS phase-shift constraint, by jointly optimizing the transmit waveform and RIS reflecting coefficients using an alternating algorithm based on the exact penalty method and manifold optimization.

When the direct paths between the BS and the targets/users are blocked in an urban environment as illustrated in Fig. 1(d), both radar sensing and communication functionalities are performed through the RIS.
The authors in \cite{Wei JCS 2022} considered this scenario, where a wideband orthogonal frequency division multiplexing (OFDM) system aims to detect one virtual LoS target in the presence of multiple clutter patches and serve multiple users with the aid of multiple RISs.
The frequency-dependent transmit beamforming and RIS phase-shifts were jointly designed to maximize the weighted radar SINR and the minimum communication SINR given a total available transmit power.
The optimization problem was solved by an alternating algorithm based on the Dinkelbach approach, majorization-minimization (MM), and alternative direction method of multipliers (ADMM).

For a more general case as illustrated in Fig. 1(e), both the LoS and non-LoS (NLoS) links have a non-negligible impact on the radar sensing performance, which results in a completely different received radar signal model compared to the cases with only LoS or NLoS link.
Specifically, the transmit waveform will reach the target via both the direct and virtual LoS links, and then be reflected back to the radar receiver (i.e., BS) through all the propagation paths.
Therefore, the target echo signals are obtained by propagating the transmit waveform through four different paths: BS$\rightarrow$target$\rightarrow$BS, BS$\rightarrow$RIS$\rightarrow$target$\rightarrow$BS, BS$\rightarrow$target$\rightarrow$RIS$\rightarrow$BS, and BS$\rightarrow$RIS$\rightarrow$target$\rightarrow$RIS$\rightarrow$BS, while only the direct path is considered in Fig. 1(c) and only the path with double reflection is considered in Fig. 1(d).
This general model allows the RIS to provide more DoFs to improve radar sensing performance by manipulating all four propagation paths.
Meanwhile, advanced algorithms are required to optimize the RIS reflecting coefficients, since the received echo signals depend on the coefficients in a complex and non-linear way, which significantly complicates the use of typical metrics such as SNR/SINR.
With this general signal model, the authors in \cite{Yan JCS 2022} assumed a simplified system model with one target and one user and investigated maximization of the radar SNR of received echo signals under the communication SNR and total power constraints.
Sophisticated derivations associated with the MM and semi-definite relaxation (SDR) methods were conducted to jointly optimize the transmit beamforming and the RIS reflecting coefficients.
The multi-user multi-input single-output (MISO) scenario was investigated in \cite{Rang EUSIPCO 2022}, where the radar receive filter was also jointly optimized to maximize the communication sum-rate under the radar SNR and total power constraints.
An alternating algorithm using fractional programming (FP), MM, and ADMM was proposed to solve the resulting problem.
A more complex scenario was considered in \cite{Rang JSTSP 2022}, where space-time adaptive processing (STAP) was utilized to combat the interference due to clutter returns.
By jointly designing the space-time transmit waveform, receive filter and the RIS reflecting coefficients using an ADMM- and MM-based algorithm, the radar SINR was maximized under the communication QoS and constant-modulus transmit waveform constraints.

In addition, physical layer security (PLS) is an important task for the cases presented in Fig. 1(f), where the target also acts as an eavesdropper attempting to extract confidential information from the high-power dual-functional signals it receives.
The authors in \cite{Mishra ICASSP 2022} proposed to guarantee the security by deploying an RIS and using artificial noise (AN) for the scenarios with only direct or double reflection paths.

\begin{figure}[t]
\centering
\includegraphics[width = \linewidth]{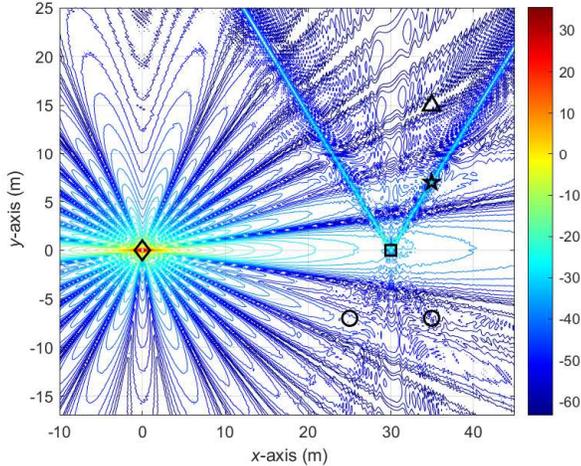}\vspace{0.1 cm}
\begin{small}\caption{Beampattern of the RIS enhanced channel. (BS: diamond, RIS: square, target: star, clutter source: triangle, users: circles, 20dBW transmit power, 10dB communication SINR, $N=256$. We assume a two-dimensional plane where the coordinates of the BS, RIS, target, clutter source, users are $(0,0)$, $(30,0)$, $(35,7)$, $(35,15)$, $(25,-7)$, $(35,-7)$, respectively.)}\end{small}
\label{fig:beampattern}
\end{figure}

\subsection{Case Study}

To reveal the potential of RIS in ISAC systems, in this section we present simulation results for an RIS-assisted DFRC system.
We assume that with the assistance of an $N$-element RIS, a 16-antenna BS simultaneously serves two single-antenna users and detects one potential point-like target in the presence of one point-like clutter source.
All the channels are LoS and follow a typical path-loss model.
The path-loss exponents for the BS-target/user/clutter source link, the RIS-target/user/clutter source link, and the BS-RIS link are set as 3.2, 2.4, and 2.2, respectively.
The other parameters are the same as that in \cite{Rang JSTSP 2022}.
The proposed algorithm in \cite{Rang JSTSP 2022} is utilized to maximize the radar SINR and satisfy the communication QoS and constant-modulus transmit power constraints.

We first illustrate the beampattern (i.e., the signal power at different locations) in Fig. 2.
It can be clearly observed that the BS generates beams towards the target, the users, and the RIS, whose passive beams further assist the BS to illuminate the target and the users.
Meanwhile, the signal power at the location of the clutter source is relatively low.

Next, the impact of the number of RIS reflecting elements is presented in Fig. 3, where the radar SNR gain provided by the received target echoes from different paths and their superposition are shown.
The scenario without an RIS is also included for comparison.
It can be observed that as $N$ increases, the performance improvement offered by the RIS is increasingly pronounced, and the echo signals from indirect paths become stronger and finally surpass that from the direct path.
In particular, the role of the LoS link dominates for relatively small $N$, where the echo signals from the direct path are the strongest and those from the double reflection are the weakest.
On the other hand, the role of the virtual LoS link created by the RIS dominates for large values of $N$ with opposite results.
These results indicate that deploying an RIS with a sufficient number of elements is competitive in improving target detection performance as well as satisfying certain communication QoS requirements.

\begin{figure}[t]
\centering
\includegraphics[width = 3.5 in]{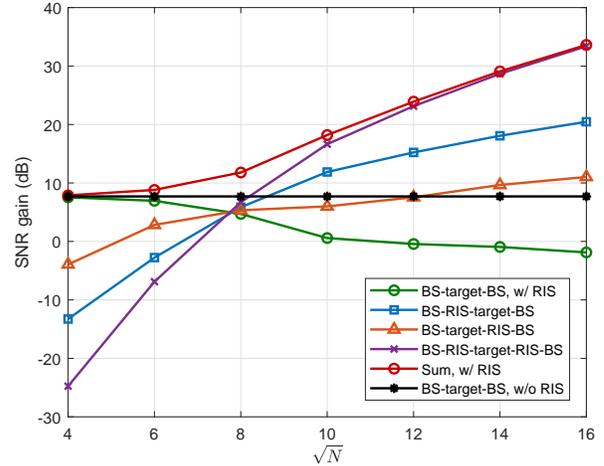}\vspace{0.1 cm}
\begin{small}\caption{Radar SNR gain of different links versus the number of RIS reflecting elements (20dBW transmit power, 10dB communication SINR).}\end{small}
\label{fig:channel gain}
\end{figure}

\begin{figure*}[!t]
\centering
\includegraphics[height = 3.6 in]{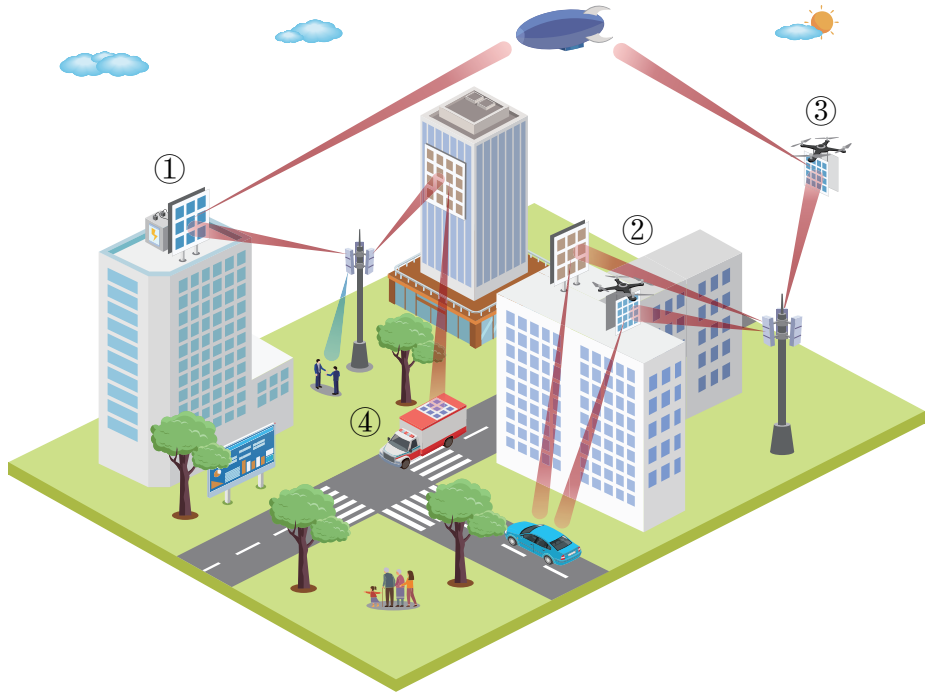}\vspace{0.2 cm}
\begin{small}\caption{Various RIS deployments in ISAC systems, in which \ding{172} an active RIS equipped with a power supply can amplify the incident signals; \ding{173} multiple RISs deployed at different locations enjoy more spatial DoFs; \ding{174} a UAV-mounted RIS has high mobility and flexibility; and \ding{175} a target-mounted RIS assists in its own detection.}\end{small}
\label{fig:various RIS deployments}\vspace{0.2 cm}
\end{figure*}

\section{Open Challenges and Future Directions}

To stimulate more practical applications of RIS in ISAC systems and capture the opportunities they provide, in this section we discuss several crucial challenges that need to be addressed and shed light on various future directions.

\subsection{Theoretical Analysis and Practical Measurement}
\subsubsection{Performance tradeoff}

Theoretical analysis is necessary to provide performance upper-bounds and to evaluate the superiority of existing solutions.
However, the theoretical upper-bounds for radar sensing have not yet been well investigated in RIS-assisted systems.
In addition, a unified upper-bound for sensing and communication performance is required to guide the designs for better radar and communication trade-offs.
It would also be meaningful to analyze the performance trade-off regions with and without RIS to reveal the potential of RIS in ISAC systems.

\subsubsection{Practical channel modeling}
As discussed in Sec. III-B, it is vital to determine the channel model for radar echoes, since different assumptions on the echo propagation paths lead to different radar signal models and algorithms.
Moreover, it is often assumed that the received target returns from all paths contribute to target detection, ignoring the propagation delays caused by reflections.
This is a slightly strict assumption considering that the propagation delay may create ghost objects and cause ranging ambiguity.
In addition, it is not clear whether there exists non-negligible interference directly reflected by the RIS or objects mounted on the RIS to the radar receiver.
One possible solution to eliminate this kind of interference is to exploit the quasi-static nature of the BS-RIS link, which generally produces interference with time-invariant statistics.
These assumptions call for experimental measurements and channel modeling based on data from system prototypes.

\subsubsection{Deployment and control of RISs}

Many existing studies are conducted assuming a given RIS deployment, i.e., where the locations, the number of reflecting elements, and the signaling to control the RISs are fixed and known in advance.
In practical heterogeneous networks, the question of where to deploy the RISs, how to determine their sizes and the signalling are also very crucial issues for achieving satisfactory performance with high efficiencies.

\begin{figure*}[!t]
\centering
\subfigure[]{
\begin{minipage}{0.32\linewidth}
\centering
\includegraphics[width = \linewidth]{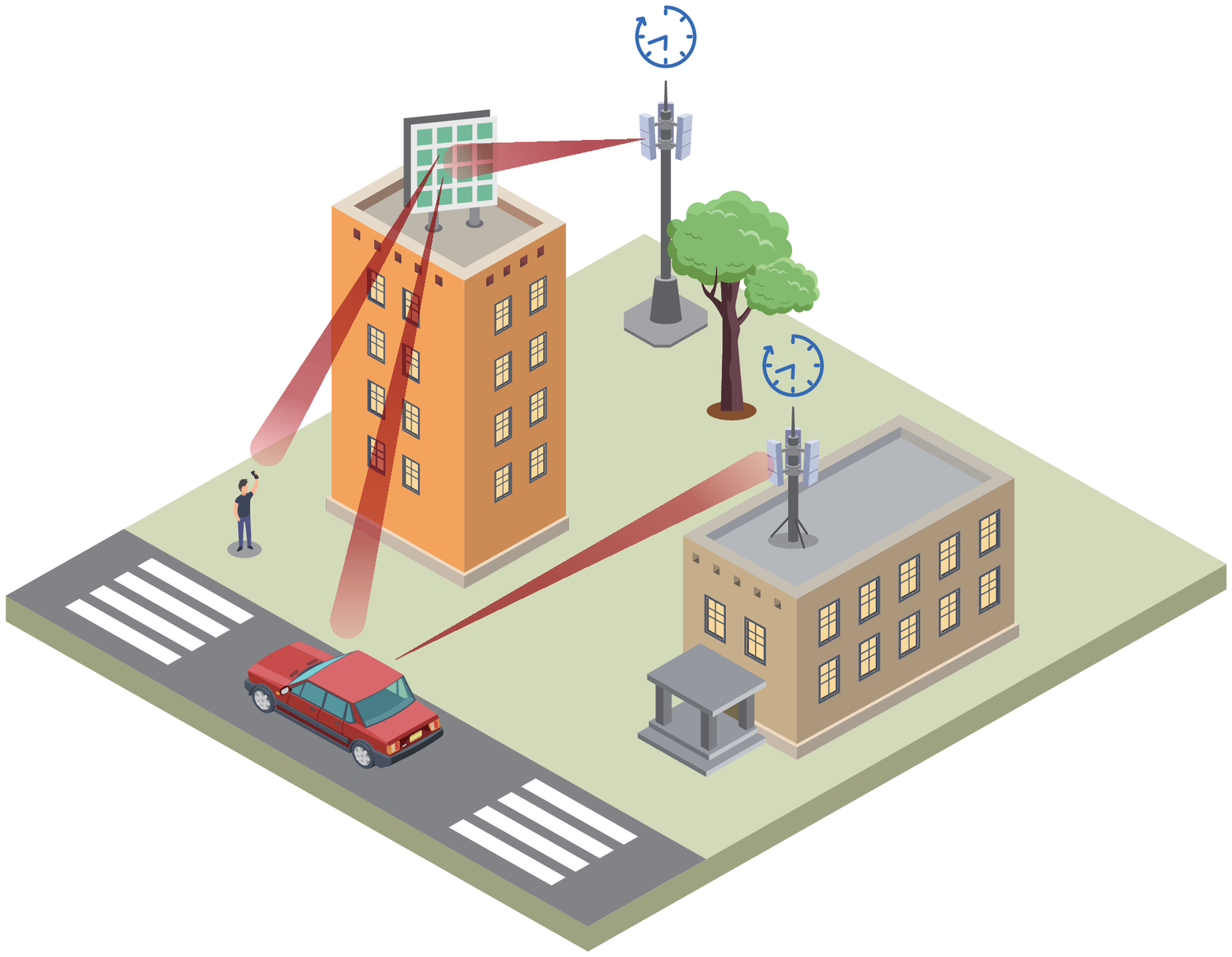}\vspace{0.3 cm}
\end{minipage}}
\subfigure[]{
\begin{minipage}{0.32\linewidth}
\centering
\includegraphics[width = \linewidth]{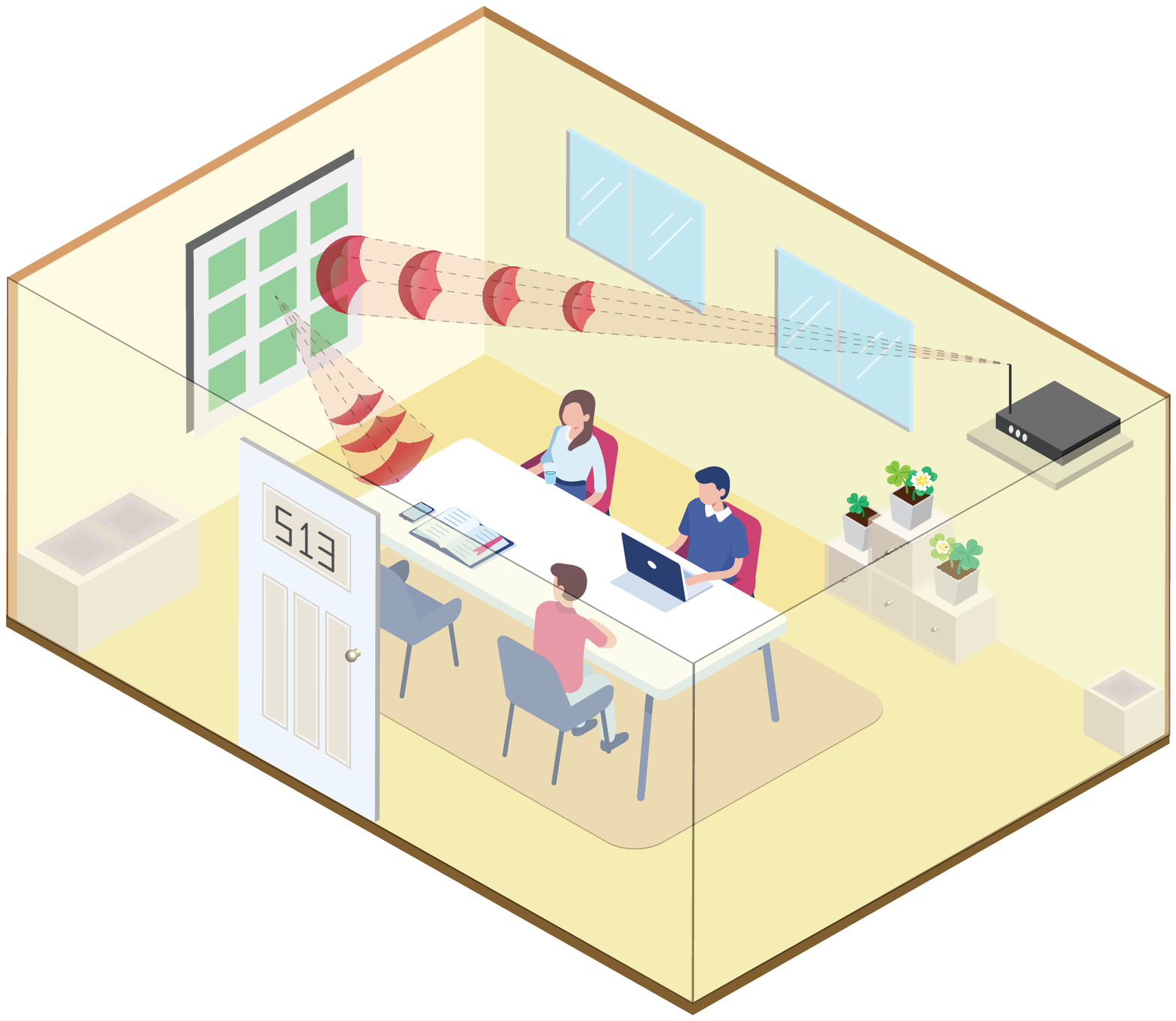}\vspace{0.3 cm}
\end{minipage}}
\subfigure[]{
\begin{minipage}{0.32\linewidth}
\centering
\includegraphics[width = \linewidth]{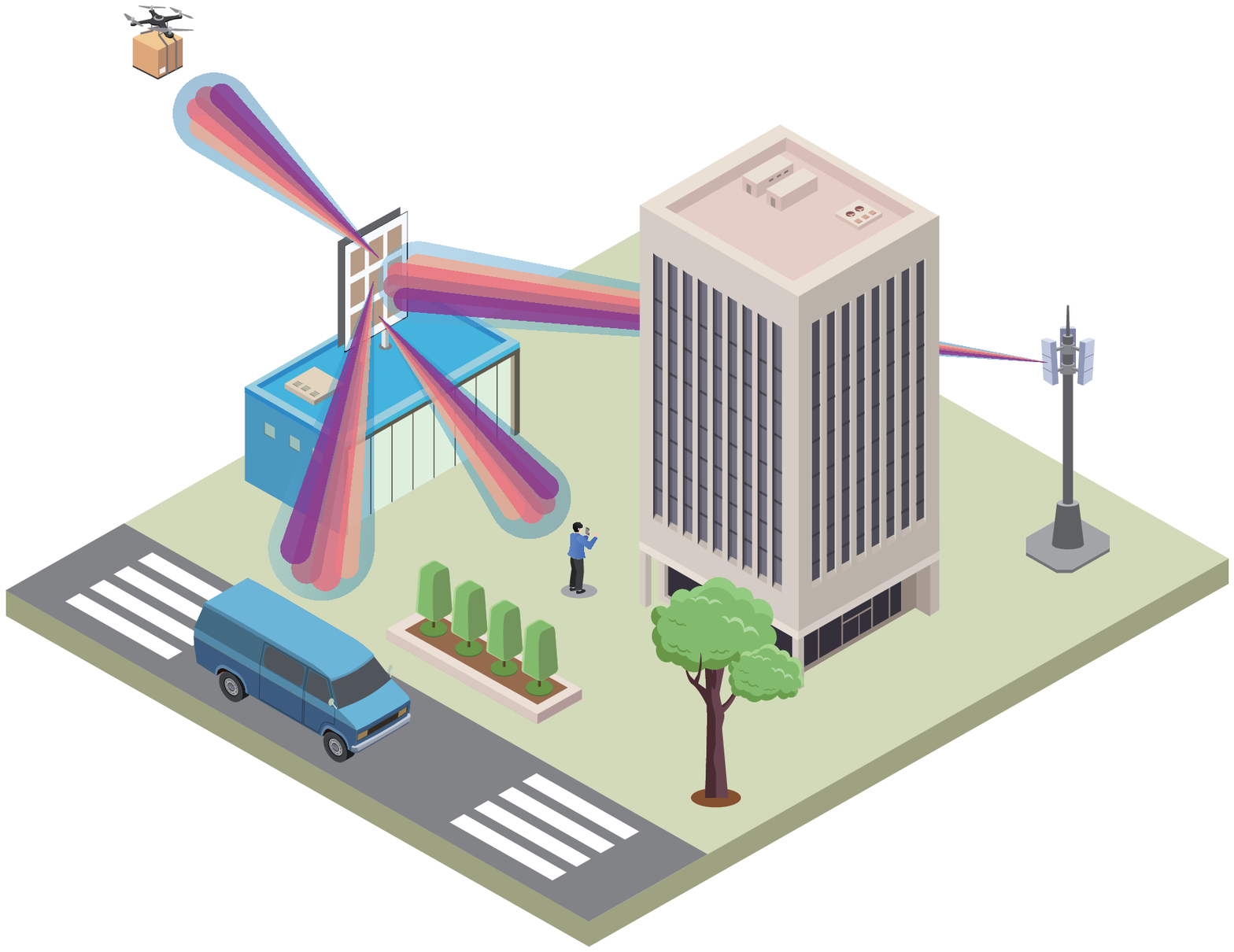}\vspace{0.3 cm}
\end{minipage}}\vspace{0.1 cm}
\begin{small}\caption{RIS-assisted ISAC scenarios. (a) RIS-assisted bi-station sensing-based ISAC system. (b) RIS-assisted near-field communication and sensing. (c) RIS-assisted wideband ISAC system. }\end{small}
\label{fig:various ISAC scenarios}\vspace{0.2 cm}
\end{figure*}

\subsection{Various RIS Deployments in ISAC Systems}

As shown in Fig. 4, individual or merged active RISs, multiple RISs, UAV-mounted and target-mounted RISs, etc., can be deployed to assist the sensing and communication functionalities in ISAC systems.

\subsubsection{Active RIS}

Active RIS equipped with reflection-type amplifiers can simultaneously tune the phase-shifts and amplify the incident signals.
Instead of using numerous passive reflecting elements, a small-sized active RIS is capable of circumventing the severe product path loss attenuation due to reflection.
While boosting the desired dual-functional signals and target returns, an active RIS also amplifies undesired noise and interference, which poses challenges to achieving satisfactory sensing and communication performance.
Deploying active RISs further causes additional power consumption and hardware complexity.
Therefore, determining the size and jointly designing the coefficients of the active RIS to provide a larger energy efficiency under certain primary sensing/communication requirements is another key problem.

\subsubsection{Multiple RISs}

Multiple RISs naturally possess additional DoFs for manipulating RF propagation environments, and their geographic diversity provides enhanced opportunities for boosting desired signals and eliminating interference.
The deployment of multiple RISs can support high-quality communication QoS and high-accuracy sensing performance in hotspots and guarantee satisfactory performance in edge areas thanks to improved passive beamforming gains.
In addition, multiple RISs offer multiple views of target echoes and increase the spatial diversity of the radar returns, which can improve detection and estimation performance.
Nevertheless, optimization of the deployment, control mechanism and protocol, as well as reflecting coefficients, must be carefully addressed to achieve the possible performance improvements.

\subsubsection{UAV-mounted RIS}

Lightweight RISs can be mounted on UAVs (e.g., conformal to the UAV wings) to provide signal propagation paths that bypass ground obstacles and create virtual LoS links to cover shadowed areas.
A UAV-mounted RIS combines the advantages of high mobility and flexibility of a UAV with the lower cost, lighter weight and lower power consumption of an RIS, making it an excellent candidate for providing reliable coverage in ISAC systems.
However, the signals reflected from the UAV may interfere with the detection of target echoes at the radar receiver; the Doppler frequency components generated by the UAV maneuvers may cause ambiguities in the resulting velocity measurements.
Therefore, the joint design of the RIS reflecting coefficients associated with the UAV trajectory and the transmit beamforming and receive processing at the BS is crucial to achieve satisfactory sensing and communication performance.

\subsubsection{Target-mounted RIS}

Deploying an RIS on the surface of a ``target'' can efficiently assist its detection by a cooperative tracking radar.
In particular, a friendly target who desires its location to be known and tracked for better beam tracking or communication may deploy its own RIS to boost the backscatter signals reflected to the radar receiver to improve the detection probability.
This is especially desirable when the target radar cross section (RCS) is small and cannot be easily detected.
To achieve this goal in a complex and changing electromagnetic environment, how to mathematically model the problem and design the reflecting coefficients and the signalling to control the RIS are key issues.

\subsection{Various RIS-assisted ISAC Scenarios}

In addition to typical RCC and DFRC scenarios, various ISAC use cases involving RIS deployments, including bi-station sensing-based ISAC, near-field communication and sensing, and wideband ISAC, deserve further investigations as presented in Fig. 5.

\subsubsection{Bi-station sensing}

Bi-static radar is popular due to its use of a passive receiver that can operate covertly and can mitigate targets intended to be cloaked, and thus motivates consideration for ISAC-based implementations.
In this context, deploying an RIS can offer two main advantages.
First, by establishing a relatively stable virtual LoS link between the transmitter and the receiver, a stronger reference signal can be guaranteed for addressing the crucial synchronization problem when the direct channel is very weak or obscured.
Second, separately deploying an RIS at the ISAC transmitter and the sensing receiver can assist in  steering the probing signal toward the target and eliminating clutter interference.

\subsubsection{Near-field communication and sensing}

Large RISs operating at high carrier frequencies have an expanded near-field region that extends well beyond that of current systems.
Moreover, it is desirable for RISs to be deployed near the transmitter or receiver in order to achieve greater passive beamforming gains.
As a result, near-field communication and sensing will become more common in RIS-assisted ISAC systems.
Unlike the planar wave model adopted for far-field transmission, electromagnetic waves in the near-field must be modeled using a spherical wave, which can not only be steered toward a specific angle but also focused on a specific distance.
Under this model, more accurate parameter estimation for sensing can be achieved, as well as improved communication capacity.
However, it is obvious that complicated channel estimation and beamforming designs are required, especially when the number of RIS reflecting elements and BS transmit/receive antennas are large.
Considering the trade-off between performance and complexity, how to intelligently  determine whether to adopt the near-field mode also deserves investigations.

\subsubsection{Wideband ISAC}

Wideband waveforms have been studied in ISAC systems since they can support high communication rates and high resolution parameter estimation.
While existing research on RIS-assisted wideband ISAC has assumed an ideal reflection model, we note that practical RISs are frequency-selective devices since each reflecting element exhibits different responses to signals at different frequencies.
System optimization without considering this characteristic will inevitably degrade communication QoS and lead to Doppler frequency ambiguity, which will affect the resulting velocity estimation.
In addition, the beam squint effect induced by large-scale RIS and the frequency-dependent steering/reflecting vectors is also worth attention.
Therefore, more accurate RIS reflection and channel models need to be established for achieving better performance in practical applications.

\subsection{Algorithm Designs}
\subsubsection{Tensor-based detection and estimation}

The signal processing at the radar receiver for detection and estimation is also very important, and has not been thoroughly investigated in most previous ISAC literature.
In MIMO radar, target detection or parameter estimation is performed by processing the radar data cube, which collects the returns in a coherent processing interval (CPI).
Taking the returns reflected by the RIS into account, the resulting high dimensional data will
lead to high computational loads if processed by conventional optimization algorithms, which suggests that alternative approaches such as tensor-based algorithms are needed.
By exploiting the inherent multi-dimensional signal structure, tensor analysis can more efficiently extract the desired target parameters from the data.
In addition, tensor decompositions offer better parameter identifiability and the possibility of closed-form solutions.

\subsubsection{AI-driven design and optimization}

Conventional optimization algorithms may not be applicable to practical RIS-assisted ISAC systems due to their prohibitively high computational complexity and idealized assumptions.
In particular, the discrete phase-shifts of the RIS and the consideration of hardware imperfections will lead to complicated non-convex optimization problems.
In addition, perfect channel state information (CSI) may not be easily obtained given the huge number of parameters that must be estimated for a large-scale RIS.
Moreover, for scenarios with multiple RISs and wideband signals, the dimension of the variables to be optimized will be extremely large.
To handle these problems, artificial intelligence (AI) is expected to provide more efficient and robust solutions.
For example, deep learning (DL) can be utilized to provide low-complexity solutions to waveform and reflection designs after off-line supervised learning.
Reinforcement learning (RL) is applicable for handling the control of the RIS or other devices in dynamic heterogeneous networks.

\section{Conclusions}

In this article, we investigate the potential of exploiting RIS technology in ISAC systems.
After briefly describing the principles of these two promising technologies, we provide an overview of the applications of RIS-assisted sensing and RIS-assisted ISAC, and present a case study to demonstrate the performance improvement introduced by the use of RIS.
We foresee that the integration of RIS and ISAC can achieve high-accuracy, wide-coverage, and ultra-reliable sensing and communication functionalities, which will be a powerful support for the vision of future 6G networks.
As this line of research is on the rise, we finally discuss some open challenges and future directions, hoping to offer a useful guide and spark research interest.

\end{document}